\documentclass[aps,prc,twocolumn,superscriptaddress,showpacs,floatfix,nofootinbib]{revtex4-1}
\usepackage{amsmath,graphicx,color}
\newcommand{\trento}{T$\mathrel{\protect\raisebox{-2.1pt}{R}}$ENTo}
\begin{document}

\title{Primordial non-Gaussianity in heavy-ion collisions}

\author{Rajeev S. Bhalerao}
\affiliation{Department of Physics, Indian Institute of Science Education and Research (IISER), Homi Bhabha Road, Pune 411008, India}
\author{Giuliano Giacalone}
\affiliation{Institut de physique th\'eorique, Universit\'e Paris Saclay, CNRS, CEA, F-91191 Gif-sur-Yvette, France} 
\author{Jean-Yves Ollitrault}
\affiliation{Institut de physique th\'eorique, Universit\'e Paris Saclay, CNRS, CEA, F-91191 Gif-sur-Yvette, France} 
\date{\today}

\begin{abstract}
  We show that experimental data on cumulants of anisotropic flow in heavy-ion collisions probe the non-Gaussian statistics of the energy density field created right after the collision.
  We carry out a perturbative expansion of the initial anisotropies of the system in terms of its density fluctuations.
  We argue that the correlation between the magnitudes of elliptic flow and triangular flow, dubbed $sc(3,2)$, is generically of the same sign and order of magnitude as the kurtosis of triangular flow in a hydrodynamic picture. 
  The experimental observation that these quantities are negative implies that the distribution of energy around a given point has positive skew.
\end{abstract}
\maketitle

\section{introduction}
Particles emitted in heavy-ion collisions display specific correlation patterns, which depend little on rapidity, but strongly on azimuthal angle~\cite{Chatrchyan:2012wg}. 
These correlations are analyzed through the observation of elliptic flow, $v_2$~\cite{Aamodt:2010pa}, and triangular flow, $v_3$~\cite{Alver:2010gr}, which are the second and third Fourier coefficients, respectively, of the azimuthal distribution of particles~\cite{Luzum:2011mm}.
In a hydrodynamic framework, the values of $v_2$ and $v_3$ originate from an almost linear response~\cite{Niemi:2012aj} to the elliptic and triangular anisotropies~\cite{Teaney:2010vd}, dubbed $\varepsilon_2$ and $\varepsilon_3$, respectively, that characterize the energy density profile of the system created in the collision process.
Since the energy density profile fluctuates event to event, $\varepsilon_2$ and $\varepsilon_3$ also fluctuate.
Their probability distribution, and consequently that of the final-state $v_n$ coefficients, carries information about the initial density fluctuations.

In this article, we use experimental data on $v_2$ and $v_3$ to gather evidence that primordial energy-density fluctuation in nucleus-nucleus collisions do not follow Gaussian statistics. 
Non-Gaussianities are generic in microscopic systems, where they appear as corrections to the central limit theorem~\cite{Alver:2008zza,Bhalerao:2011bp,Yan:2013laa}. 
Unlike the situation in the early Universe, where primordial non-Gaussianities are compatible with zero~\cite{Ade:2013ydc}, it is natural to expect that they are sizable in the early stages of heavy-ion collisions. 

Since the relation between the initial anisotropies and the initial density is not linear, non-Gaussian fluctuations of $\varepsilon_2$ and $\varepsilon_3$ are nontrivially related to non-Gaussian fluctuations of the initial energy density. 
Our aim in this paper is to derive expressions relating them. 
We show that the correlation between the magnitudes of $v_2$ and $v_3$, as measured by the symmetric cumulant $SC(3,2)$~\cite{Bilandzic:2013kga,ALICE:2016kpq,STAR:2018fpo}, and the kurtosis of the distribution of $v_3$~\cite{Abbasi:2017ajp} as measured by $v_3\{4\}$~\cite{ALICE:2011ab}, have a common origin. 
Their negative values can be ascribed to the positive skewness of the distribution of energy in a given transverse area. 
We also show that, paradoxically, the skewness of elliptic flow fluctuations~\cite{Giacalone:2016eyu,Sirunyan:2017fts,Acharya:2018lmh} is likely to be less sensitive to the skewness of density fluctuations. 

In order to relate the fluctuations of the initial anistotropies to the fluctuations of the initial density profile, we carry out a systematic perturbative expansion in powers of the density fluctuations~\cite{Blaizot:2014nia,Gronqvist:2016hym}, which is introduced in Sec.~\ref{s:pert}. 
This is a general valid approach for large systems such as nucleus-nucleus collisions.
Small systems (proton-nucleus and proton-proton collisions~\cite{Albacete:2017ajt,Sirunyan:2017uyl,Aaboud:2018syf,Acharya:2019vdf}) are not considered here. 
For simplicity, the calculations in the main body of this article are carried out for central collisions (i.e., at zero impact parameter).
As we shall argue, the geometry of non-central collisions only matters when studying the fluctuations of $\varepsilon_2$, whose discussion is relegated to Appendix~\ref{s:noncentral}. 
In Sec.~\ref{s:cumulants}, we recall the definitions of cumulants of $\varepsilon_n$, and the definitions of cumulants of the initial energy density field.
In Sec.~\ref{s:calcul}, we evaluate the cumulants of order 4 of the distribution of $\varepsilon_n$, including the \textit{mixed} cumulant $SC(3,2)$, to leading order in the perturbative expansion.
In Sec.~\ref{s:standardized}, we test these perturbative results using Monte Carlo calculations. 
We define several measures of primordial non-Gaussianity which allow to compare models of initial conditions with experimental data, and we discuss the implications of current measurements.

\section{Perturbative expansion of $\varepsilon_n$}
\label{s:pert}

We first introduce a few notations. 
We denote the initial energy density profile of an event at mid-rapidity by $\rho({\bf s})$, where ${\bf s}$ labels a point in the transverse plane.\footnote{We shall not deal with the longitudinal dynamics of the system, which is not relevant for our discussion.}
We are interested in the event-by-event fluctuations of $\rho({\bf s})$ in a situation where macroscopic quantities (typically, the centers of colliding nuclei) are fixed.
We denote the statistical average of $\rho({\bf s})$ by $\langle\rho({\bf s})\rangle$, and the the event-to-event fluctuation by $\delta\rho({\bf s})\equiv\rho({\bf s})-\langle\rho({\bf s})\rangle$. 
We choose a coordinate frame where the origin of the transverse plane lies at the center of the average energy density, so that
\begin{equation}
\label{deforigin} 
\int_{\bf s}{\bf s}\langle\rho({\bf s})\rangle=0,
\end{equation}
where we use the short hand $\int_{{\bf s}}=\int {\rm d}x{\rm d}y$ for the integration over the transverse  plane.
Due to fluctuations, the center of the distribution $\rho({\bf s})$, which we denote by ${\bf s}_0$, fluctuates event to event.
It is defined by:
\begin{equation}
\label{defs0}
  {\bf s}_0\equiv\frac{\int_{\bf s}{\bf s}\rho({\bf s})}{\int_{\bf s}\rho({\bf s})}=\frac{\int_{\bf s}{\bf s}\delta\rho({\bf s})}{\int_{\bf s}\rho({\bf s})}.
  \end{equation}
The initial eccentricity $\varepsilon_2$ and the initial triangularity $\varepsilon_3$ are then defined by
\begin{equation}
  \label{defepsn2}
  \varepsilon_n=\frac{\int_{\bf s}({\bf s}-{\bf s}_0)^n \rho({\bf s})}
  {\int_{\bf s}|{\bf s}-{\bf s}_0|^n \rho({\bf s})}, 
\end{equation}
where we have used the complex notation ${\bf s}=x+iy$, and the recentering correction ${\bf s}_0$ ensures that anisotropies are evaluated in a centered frame~\cite{Teaney:2010vd,Alver:2006wh}.
Under the transformation ${\bf s}\to{\bf s}e^{i\alpha}$, $\varepsilon_2$ and $\varepsilon_3$ are multiplied by $e^{2i\alpha}$ and $e^{3i\alpha}$, respectively, thus $\varepsilon_n$ is a measure of the deformation of $\rho({\bf s})$ in the $n^{\rm th}$ Fourier harmonic. 

We now carry out a perturbative expansion of $\varepsilon_n$. 
We decompose $\rho=\langle \rho\rangle+\delta\rho$, and expand the right hand side of Eq.~(\ref{defepsn2}) in powers of $\delta\rho$,\footnote{Note that even though $\delta\rho({\bf s})$ may be of the same order, or even larger than the average density $\langle\rho({\bf s})\rangle$ locally, its contribution to $\varepsilon_n$ remains small for a large system~\cite{Blaizot:2014nia}.} keeping in mind that ${\bf s}_0$ is itself of order $\delta\rho$.
For simplicity, we consider a central collision where the average density $\langle\rho({\bf s})\rangle$ is azimuthally symmetric, so that $\varepsilon_n$ is solely due to fluctuations.
The noncentral case is discussed in Appendix~\ref{s:noncentral}.
In central collisions, there is no mean anisotropy, and the dominant contribution to $\varepsilon_n$ is of order $\delta\rho$. 
In order to evaluate the non-Gaussian fluctuations to leading order, we shall need to include the next-to-leading correction, of order $(\delta\rho)^2$. 
We show in Appendix~\ref{s:nnlo} that the next-to-next-to-leading correction, of order $(\delta\rho)^3$, does not contribute to the quantities calculated in this article. 

In order to achieve compact expressions, we introduce the following shorthand notations, for any function $f({\bf s})$~\cite{Gronqvist:2016hym}:
\begin{eqnarray}
\label{notation}
\delta f &\equiv&\frac{1}{\langle E\rangle}\int_{\bf s} f({\bf s})\delta\rho({\bf s})\cr
\langle f\rangle &\equiv&\frac{1}{\langle E\rangle}\int_{\bf s} f({\bf s})\langle\rho({\bf s})\rangle,
\end{eqnarray}
where $\langle E\rangle$ is the average energy per longitudinal length:
\begin{equation}
\label{defeav}
\langle E\rangle=\int_{\bf s} \langle\rho({\bf s})\rangle.
\end{equation}
Thus $\langle f\rangle$ is the average value of $f({\bf s})$ with the weight $\langle\rho({\bf s})\rangle$.
With these notations, the center of the distribution is ${\bf s}_0\simeq \delta {\bf s}$, where we have made the approximation $\rho({\bf s})\simeq\langle\rho({\bf s})\rangle$ in the denominator of Eq.~(\ref{defs0}), which amounts to expanding ${\bf s}_0$ to leading order in $\delta\rho$. 
Expanding Eq.~(\ref{defepsn2}) to second order in $\delta\rho$, we then obtain:
\begin{eqnarray}
  \label{perteps}
  \varepsilon_2&=&
\frac{\delta {\bf s}^2}{\langle |{\bf s}|^2\rangle}
  -\frac{(\delta |{\bf s}|^2)(\delta {\bf s}^2)}{\langle |{\bf s}|^2\rangle^2}
  -\frac{(\delta {\bf s})^2}{\langle |{\bf s}|^2\rangle},\cr
  \varepsilon_3&=&
\frac{\delta {\bf s}^3}{\langle |{\bf s}|^3\rangle}
  -\frac{(\delta |{\bf s}|^3)(\delta {\bf s}^3)}{\langle |{\bf s}|^3\rangle^2}
  -3\frac{(\delta {\bf s}^2)(\delta {\bf s})}{\langle |{\bf s}|^3\rangle}.
\end{eqnarray}
In each line, there is one term of order $\delta\rho$, which is the dominant contribution to the anisotropy, and was the only term kept in \cite{Blaizot:2014nia}.
In addition, there are two next-to-leading corrections of order $(\delta\rho)^2$, which originate from the non-linear relation between $\varepsilon_n$ and $\delta\rho$~\cite{Gronqvist:2016hym}.
The first correction is the effect of the fluctuation in the fireball size (denominator of Eq.~(\ref{defepsn2})), while the last term originates from the recentering correction. 
Note that the recentering correction was neglected in~\cite{Gronqvist:2016hym}.
We retain it in this paper because, as will be shown below, it plays a crucial role for the symmetric cumulant $SC(3,2)$.

\section{Definitions of cumulants}
\label{s:cumulants}

\subsection{Cumulants of initial anisotropies}

Anisotropic flow is not measured on an event-by-event basis.
Experimentally-measured quantities are multi-particle correlations averaged over events, which yield average values of products of $v_n$~\cite{Bhalerao:2014xra}.
If $v_n$ is proportional to $\varepsilon_n$ in every event~\cite{Gardim:2011xv}, then, the relevant quantities for phenomenology are average values of products of $\varepsilon_n$, i.e., moments.
Since $\varepsilon_n$ is a complex number whose phase is uniformly distributed by azimuthal symmetry, moments which are not invariant under rotations, such as $\langle \varepsilon_n\rangle$ or $\langle (\varepsilon_n)^2\rangle$, vanish by symmetry. 
The lowest-order non-trivial moment is the rms value of $\varepsilon_n$, denoted by $c_n\{2\}$: 
\begin{equation}
  \label{defepsc2}
c_n\{2\}\equiv\langle \varepsilon_n\varepsilon_n^*\rangle, 
\end{equation}
where $\varepsilon_n^*$ denotes the complex conjugate, and angular brackets an average over events.\footnote{Note that we use the same angular brackets to denote an average over events, or an average value taken with the mean density profile, as in Eq.~(\ref{notation}). There should be no confusion depending on the context.}

Moments of order $\ge 4$ (note that all odd moments vanish by azimuthal symmetry) are usually combined with lower-order moments to form cumulants.
Cumulants were originally introduced to suppress nonflow correlations in experimental analyses~\cite{Borghini:2000sa}, but they turn out to be useful observables also for studying flow fluctuations~\cite{Voloshin:2007pc,Yan:2013laa}.
In this article, we study cumulants of order 4, also called kurtosises, which are obtained from moments of order 4 by subtracting all pairwise correlations.
There are cumulants involving only one harmonic $n$, denoted by
$c_n\{4\}$~\cite{Borghini:2001vi}:
\begin{eqnarray}
\label{defcn4}
  c_n\{4\}
  &\equiv& \langle \varepsilon_n \varepsilon_n\varepsilon_n^*\varepsilon_n^*\rangle
-2\langle\varepsilon_n \varepsilon_n^*\rangle\langle\varepsilon_n\varepsilon_n^*\rangle
-\langle\varepsilon_n \varepsilon_n\rangle\langle\varepsilon_n^*\varepsilon_n^*\rangle\cr
  &=& \langle \varepsilon_n \varepsilon_n\varepsilon_n^*\varepsilon_n^*\rangle
-2\langle\varepsilon_n \varepsilon_n^*\rangle\langle\varepsilon_n\varepsilon_n^*\rangle.
\end{eqnarray}
In the second equality, we have used azimuthal symmetry to simplify the expression. 
There is also a cumulant involving both $\varepsilon_2$ and $\varepsilon_3$, called symmetric cumulant, and denoted by $SC(3,2)$.\footnote{This quantity was denoted by $SC(3,2)_\varepsilon$ in \cite{ALICE:2016kpq} to avoid confusion with the measured quantity, which involves $v_n$ rather than $\varepsilon_n$.}
It is defined in a similar way~\cite{Bilandzic:2013kga}:
\begin{eqnarray}
  \label{defsc32}
 SC(3,2)&\equiv& \langle \varepsilon_2 \varepsilon_3\varepsilon_2^*\varepsilon_3^*\rangle
-\langle\varepsilon_2 \varepsilon_3\rangle\langle\varepsilon_2^*\varepsilon_3^*\rangle\cr
&& -\langle\varepsilon_2 \varepsilon_2^*\rangle\langle\varepsilon_3\varepsilon_3^*\rangle
-\langle\varepsilon_2 \varepsilon_3^*\rangle\langle\varepsilon_3\varepsilon_2^*\rangle \cr
&=& \langle \varepsilon_2 \varepsilon_3\varepsilon_2^*\varepsilon_3^*\rangle
-\langle\varepsilon_2 \varepsilon_2^*\rangle\langle\varepsilon_3\varepsilon_3^*\rangle,
\end{eqnarray}
where, in the last equality, we have again eliminated the terms that vanish by symmetry. In this article, we refer to $SC(3,2)$ as to the ``mixed kurtosis'', because it involves two different Fourier harmonics. 

\subsection{Cumulants of initial density}

If one replaces $\varepsilon_n$ by its perturbative expansion, 
Eq.~(\ref{perteps}), moments and cumulants involve 
average values of products of $\delta\rho$, that is, correlation functions of the field $\rho({\bf s})$, which we now define.

The 2-point function, $C_2({\bf s}_1,{\bf s}_2)$, characterizes the variance of the fluctuations. It is defined by:
\begin{equation}
 \label{defC2}
 C_2({\bf s}_1,{\bf s}_2)  \equiv \langle\delta\rho({\bf s}_1)\delta\rho({\bf s}_2)\rangle,
\end{equation}
where we used $\langle \delta\rho ({\bf s}) \rangle=0$.
Similarly, the connected 3-point function, which characterizes the skewness of density fluctuations, is defined by:
\begin{equation}
\label{defC3}
  C_3({\bf s}_1,{\bf s}_2,{\bf s}_3)  \equiv \langle\delta\rho({\bf s}_1)\delta\rho({\bf s}_2)\delta\rho({\bf s}_3)\rangle.
\end{equation}
In this article, we shall also need the connected 4-point function, or kurtosis of density fluctuations, which is defined by subtracting out all pairwise correlations:
\begin{eqnarray}
\label{defC4}
C_4({\bf s}_1,{\bf s}_2,{\bf s}_3,{\bf s}_4)  &\equiv& \langle\delta\rho({\bf s}_1)\delta\rho({\bf s}_2)\delta\rho({\bf s}_3)\delta\rho({\bf s}_4)\rangle\cr
&-&\langle\delta\rho({\bf s}_1)\delta\rho({\bf s}_2)\rangle\langle\delta\rho({\bf s}_3)\delta\rho({\bf s}_4)\rangle\cr
&-&\langle\delta\rho({\bf s}_1)\delta\rho({\bf s}_3)\rangle\langle\delta\rho({\bf s}_2)\delta\rho({\bf s}_4)\rangle\cr
&-&\langle\delta\rho({\bf s}_1)\delta\rho({\bf s}_4)\rangle\langle\delta\rho({\bf s}_2)\delta\rho({\bf s}_3)\rangle.
\end{eqnarray}
For Gaussian density fluctuations~\cite{Floerchinger:2014fta}, $C_3$ and $C_4$ vanish. 
However, the positivity condition $\rho(x)\ge 0$ naturally generates positive cumulants to all orders~\cite{Gronqvist:2016hym}.
One typically expects that the functions $C_n$ are short ranged (they vanish unless all the arguments are close to one another) and positive.
However, long-range correlations can be induced by energy conservation, i.e., if one requires that all events have the exact same energy~\cite{Blaizot:2014nia,Gronqvist:2016hym}.

\subsection{$n$-point averages and orders of magnitude}
\label{s:npointav}

In this article, we shall need to evaluate weighted products of $\delta\rho$, averaged over events, up to order $(\delta\rho)^6$. We now explain how such products are evaluated. 
For any function $f$, the one-point average $\langle \delta f\rangle$ vanishes by definition of $\delta\rho$.
The first non-trivial average is the two-point average, which is obtained by multiplying Eq.~(\ref{defC2}) with functions of transverse coordinates and integrating over the transverse plane:
\begin{equation}
\label{2pointav}
\langle\delta f\delta g\rangle
=\frac{1}{\langle E\rangle^2}\int_{{\bf s}_1,{\bf s}_2} f({\bf s}_1)g({\bf s}_2)
C_2({\bf s}_1,{\bf s}_2).
\end{equation}
The three-point average is obtained in a similar way using Eq.~(\ref{defC3}):
\begin{equation}
\label{3pointav}
\langle\delta f\delta g\delta h\rangle=
\frac{1}{\langle E\rangle^3}\int_{{\bf s}_1,{\bf s}_2,{\bf s}_3} f({\bf s}_1)g({\bf s}_2)h({\bf s}_3) C_3({\bf s}_1,{\bf s}_2,{\bf s}_3).
\end{equation}
Four-point averages are decomposed according to Wick's theorem into products of two-point averages and a connected part, which we denote by the subscript $c$:
\begin{eqnarray}
\label{4pointav}
\langle\delta f\delta g\delta h\delta k\rangle_c&\equiv&
\langle\delta f\delta g\delta h\delta k\rangle
-\langle\delta f\delta g\rangle\langle\delta h\delta k\rangle\cr
&&-\langle\delta f\delta h\rangle\langle\delta g\delta k\rangle-\langle\delta f\delta k\rangle\langle\delta g\delta h\rangle.
\end{eqnarray}
With this notation, Eqs.~(\ref{defC4}) and (\ref{2pointav}) give immediately:
\begin{widetext}
\begin{equation}
\label{4pointavc}
\langle\delta f\delta g\delta h\delta k\rangle_c=
\frac{1}{\langle E\rangle^4}\int_{{\bf s}_1,{\bf s}_2,{\bf s}_3,{\bf s}_4} f({\bf s}_1)g({\bf s}_2)h({\bf s}_3)k({\bf s}_4) C_4({\bf s}_1,{\bf s}_2,{\bf s}_3,{\bf s}_4).
\end{equation}
\end{widetext}
For short-range correlations, one can make the approximation that the weighting functions $f,g,h,k$ in Eqs.~(\ref{2pointav}), (\ref{3pointav}) and (\ref{4pointavc}) are evaluated at the same point~\cite{Gronqvist:2016hym}, and integrate $C_n$ over relative positions~\cite{Giacalone:2019kgg}.
Thus $\langle\delta f\delta g\rangle$ is just a weighted integral of $f({\bf s})g({\bf s})$, with a positive weight, and so on. 


Let us now discuss orders of magnitude. If the weighting function $f$ is of order unity, then, $\delta f$  is typically much smaller than unity.
Hence, the $n$-point averages are strongly ordered as a function of $n$.
One must however distinguish odd and even moments. 
If $n$ is even, $(\delta f)^n$ is typically of the same order of magnitude as its average value over events  $\langle (\delta f)^{n}\rangle$.
For odd moments, there are cancellations upon averaging over events, so that  $\langle (\delta f)^{2n-1}\rangle$ is of the same order as $\langle (\delta f)^{2n}\rangle$ for $n\ge 2$. 
For short-range correlations, the {\it connected\/} part $\langle (\delta f)^{n}\rangle_c$ is 
much smaller than $\langle (\delta f)^{n}\rangle$ for $n\ge 4$.

In this article, we shall need to evaluate 5-point and 6-point averages, but only to leading order. 
These leading-order contributions are the disconnected terms given by the Wick decomposition.
The 5-point average is decomposed as a sum of products of 2-point averages and 3-point averages:
\begin{eqnarray}
\label{5pointav}
\langle\delta f\delta g\delta h\delta k\delta l\rangle&=&
\langle\delta f\delta g\rangle\langle\delta h\delta k\delta l\rangle\cr
&&+{\rm permutations}\ (10\ {\rm terms}),
\end{eqnarray}
while the dominant contribution to a 6-point average is from products of 2-point averages:
\begin{eqnarray}
\label{6pointav}
\langle\delta f\delta g\delta h\delta k\delta l\delta m\rangle&=&
\langle\delta f\delta g\rangle\langle\delta h\delta k\rangle\langle\delta l\delta m\rangle\cr
&&+{\rm permutations}\ (15\ {\rm terms}).
\end{eqnarray}

\section{Perturbative evaluation of cumulants}
\label{s:calcul}

\subsection{General expressions}
\label{s:general}

The perturbative expression of $c_n\{2\}$ in terms of the two-point function was derived in \cite{Blaizot:2014nia} to leading order $(\delta\rho)^2$.
It is obtained by inserting the first term in the right-hand side of Eqs.~(\ref{perteps}) into Eq.~(\ref{defepsc2}):
\begin{equation}
  \label{oldresult}
c_n\{2\}=\frac{\langle\delta {\bf s}^n\delta {\bf s}^{{\bf *}n}\rangle}{\langle |{\bf s}|^n\rangle^2}. 
\end{equation}
In this article, we shall evaluate cumulants of order 4 to the first non-trivial order. 

We start with the leading contribution to $\varepsilon_n$, corresponding to the first term in the 
right-hand side of Eq.~(\ref{perteps}), which is proportional to $\delta\rho$.
One naively expects that its contribution to cumulants of order 4 is of order $(\delta\rho)^4$.
However, this dominant contribution vanishes after averaging over events and subtracting the pairwise correlations in Eqs.~(\ref{defcn4}) and (\ref{defsc32}). 
Only the subleading connected part, defined by Eq.~(\ref{4pointav}), remains:
\begin{eqnarray}
\label{c4A}
  c_n\{4\}_K&=&\frac{\left\langle\delta {\bf s}^n\delta {\bf s}^n\delta {\bf s}^{{\bf *}n}\delta {\bf s}^{{\bf *}n}\right\rangle_c}{\langle |{\bf s}|^n\rangle^4}\cr
SC(3,2)_K&=&\frac{\left\langle\delta {\bf s}^2\delta {\bf s}^3\delta {\bf s}^{{\bf *}2}\delta {\bf s}^{{\bf *}3}\right\rangle_c}{\langle |{\bf s}|^2\rangle^2\langle |{\bf s}|^3\rangle^2}.
\end{eqnarray}
The numerator in these equations is proportional to the kurtosis of the density field $C_4$, according to Eq.~(\ref{4pointavc}), which is the reason why we label it with the subscript $K$.
Note that $c_n\{4\}_K$ and  $SC(3,2)_K$ are positive for short-range correlations.
Indeed, for the reasons explained in Sec.~\ref{s:npointav}, the numerators in the right-hand side of Eq.~(\ref{c4A}) are integrals of $|{\bf s}|^{4n}$ and $|{\bf s}|^{10}$, respectively, with a positive weight.

If the anisotropies depended linearly on the initial density $\rho$, the kurtosises of the initial anisotropy would be proportional to $C_4$, the kurtosis of density fluctuations.
However, the nonlinearity, which results in the second-order terms in Eq.~(\ref{perteps}), generates other contributions to the kurtosises which are of the same order as $c_n\{4\}_K$ and  $SC(3,2)_K$. 
These contributions originate from terms of order $(\delta\rho)^5$ and $(\delta\rho)^6$ in the expansion of the product of the four $\varepsilon_n$ factors in Eqs.~(\ref{defcn4}) and (\ref{defsc32}). 

We now evaluate the contribution of order $(\delta\rho)^5$, which is obtained by keeping the correction of order $(\delta\rho)^2$ for just one of the four factors, and only the terms of order $\delta\rho$ for the other three factors. 
The corresponding contribution to the product of $\varepsilon_n$ is of the type $\delta_2\delta_2\delta_1\delta_1\delta_1$, where we use the shorthand notation $\delta_1$ for the leading contribution to $\varepsilon_n$ or $\varepsilon_n^*$, and $\delta_2\delta_2$ for one of the two subleading contributions. 
When averaging over events, this 5-point average is split according to Wick's theorem, Eq.~(\ref{5pointav}).
The contractions of the type  $\langle\delta_2\delta_2\rangle\langle\delta_1\delta_1\delta_1\rangle$ or $\langle\delta_2\delta_2\delta_1\rangle\langle\delta_1\delta_1\rangle$ cancel out when subtracting the disconnected parts in Eqs.~(\ref{defcn4}) and (\ref{defsc32}).
Only the contractions of the type $\langle\delta_2\delta_1\rangle\langle\delta_2\delta_1\delta_1\rangle$ remain.
Using azimuthal symmetry to simplify the expressions, one obtains:
\begin{eqnarray}
\label{c4B}
c_n\{4\}_S&=&-8\frac{\left\langle\delta {\bf s}^n\delta {\bf s}^{{\bf *}n}\right\rangle
\left\langle\delta|{\bf s}|^n\delta {\bf s}^n\delta {\bf s}^{{\bf *}n}\right\rangle}
{\langle |{\bf s}|^n\rangle^5},
\cr
SC(3,2)_S&=&-2\frac{\left\langle\delta {\bf s}^3\delta {\bf s}^{{\bf *}3}\right\rangle
\left\langle\delta|{\bf s}|^3\delta {\bf s}^2\delta {\bf s}^{{\bf *}2}\right\rangle}
{\langle |{\bf s}|^2\rangle^2\langle |{\bf s}|^3\rangle^3}
\cr
&&-2\frac{\left\langle\delta {\bf s}^2\delta {\bf s}^{{\bf *}2}\right\rangle
\left\langle\delta|{\bf s}|^2\delta {\bf s}^3\delta {\bf s}^{{\bf *}3}\right\rangle}
{\langle |{\bf s}|^2\rangle^3\langle |{\bf s}|^3\rangle^2}
\cr
&&-6\frac{\left\langle\delta {\bf s}^2\delta {\bf s}^{{\bf *}2}\right\rangle
\left\langle\delta{\bf s}\delta {\bf s}^2\delta {\bf s}^{{\bf *}3}\right\rangle}
{\langle |{\bf s}|^2\rangle^2\langle |{\bf s}|^3\rangle^2},
\end{eqnarray}
where we use the subscript $S$ because this contribution involves three-point averages, which are proportional to the skewness of the density field $C_3$, according to Eq.~(\ref{3pointav}). 
The only contribution to $c_n\{4\}_S$ is from the second term in the right-hand side of Eq.~(\ref{perteps}), i.e., the fluctuation in the size.
The recentering correction does not contribute to $c_n\{4\}_S$, but it is the origin of the last term in $SC(3,2)_S$. 
Note that $c_n\{4\}_S$ and $SC(3,2)_S$ are always negative for short-range correlations, contrary to $c_n\{4\}_K$ and $SC(3,2)_K$ which are positive. 

Finally, we consider contributions of order $(\delta\rho)^6$, whose average over events are evaluated using Eq.~(\ref{6pointav}). 
Such contributions can arise in two different ways.
The first possibility is to have a correction of order $(\delta\rho)^3$ to one of the four $\varepsilon_n$ factors, so that the product is of the type $(\delta_3\delta_3\delta_3)\delta_1\delta_1\delta_1$. 
We show in Appendix~\ref{s:nnlo} that these terms do not contribute to the kurtosises. 
The other contributions of order $(\delta\rho)^6$ arise from having a correction of order $(\delta\rho)^2$ to two of the four $\varepsilon_n$ factors, so that the product is of the type $(\delta_2\delta_2)(\delta'_2\delta'_2)\delta_1\delta_1$. 
The contractions 
$\langle\delta_2\delta_2\rangle\langle\delta'_2\delta'_2\rangle\langle\delta_1\delta_1\rangle$
or
$\langle\delta_2\delta'_2\rangle\langle\delta_2\delta'_2\rangle\langle\delta_1\delta_1\rangle$
cancel out in the subtraction of  Eqs.~(\ref{defcn4}) and (\ref{defsc32}), while the contractions 
$\langle\delta_1\delta'_2\rangle\langle\delta_1\delta'_2\rangle\langle\delta_2\delta_2\rangle$
or
$\langle\delta_1\delta_2\rangle\langle\delta_1\delta_2\rangle\langle\delta'_2\delta'_2\rangle$
vanish by azimuthal symmetry.
Eventually, only the contractions 
$\langle\delta_1\delta_2\rangle\langle\delta_1\delta'_2\rangle\langle\delta_2\delta'_2\rangle$
contribute to the cumulant.
One obtains 
\begin{eqnarray}
\label{c4C}
c_n\{4\}_V&=&8\frac{\left\langle\delta {\bf s}^n\delta {\bf s}^{{\bf *}n}\right\rangle^2
\left\langle\delta|{\bf s}|^n\delta |{\bf s}|^n\right\rangle}
{\langle |{\bf s}|^n\rangle^6},
\cr
SC(3,2)_V&=&4\frac{\left\langle\delta {\bf s}^2\delta {\bf s}^{{\bf *}2}\right\rangle\left\langle\delta {\bf s}^3\delta {\bf s}^{{\bf *}3}\right\rangle
\left\langle\delta|{\bf s}|^2\delta |{\bf s}|^3\right\rangle}
{\langle |{\bf s}|^2\rangle^3\langle |{\bf s}|^3\rangle^3}
\cr
&&+
9\frac{\left\langle\delta {\bf s}^2\delta {\bf s}^{{\bf *}2}\right\rangle^2\left\langle\delta {\bf s}\delta {\bf s^*}\right\rangle}
{\langle |{\bf s}|^2\rangle^2\langle |{\bf s}|^3\rangle^2},
\end{eqnarray}
where the subscript $V$ means that this contribution only involves two-point averages (\ref{2pointav}), i.e., the variance $C_2$ of the density field. 
The only contribution to $c_n\{4\}_V$ comes from the size fluctuation (second term in the right-hand side of Eqs.~(\ref{perteps})).
The first contribution to $SC(3,2)_V$ is also from the size fluctuation, while the second is from the recentering correction to $\varepsilon_3$ (last term in the second line of Eq.~(\ref{perteps})).
Both $c_n\{4\}_V$ and $SC(3,2)_V$ are positive. 

The full expressions of the cumulants to leading order in the perturbative expansion are obtained by summing the contributions of Eqs.~(\ref{c4A}), (\ref{c4B}), (\ref{c4C}):
\begin{eqnarray}
  \label{fullc4}
  c_n\{4\}&=&c_n\{4\}_K+c_n\{4\}_S+c_n\{4\}_V\cr
  SC(3,2)&=&  SC(3,2)_K+ SC(3,2)_S+ SC(3,2)_V.
\end{eqnarray}
Note that the recentering correction does not contribute to $c_n\{4\}$ (which is the reason why it was neglected in~\cite{Gronqvist:2016hym}), but it contributes to $SC(3,2)$.

As we shall illustrate in the specific case of identical pointlike sources~\cite{Bhalerao:2006tp}, $V$ and $K$ contributions are positive, while the $S$ contribution is negative.
Therefore, the experimental observation that both $c_3\{4\}$ and $SC(3,2)$ are negative implies that the $S$-type contribution, which is induced by the 3-point function of the density field, dominates.
We come back to this point in Sec.~\ref{s:expdata}. 

\subsection{Identical pointlike sources}
\label{s:pointlike}

We now simplify the results of Sec.~\ref{s:general} in the specific case where the energy density is a sum of Dirac peaks~\cite{Gronqvist:2016hym}:
\begin{equation}
\rho({\bf s})=\sum_{j=1}^N\delta({\bf s}-{\bf s}_j), 
\end{equation}
where $N$ is a fixed number, and ${\bf s}_j$ are independent variables, whose probability distribution is defined by the average density $\langle\rho({\bf s})\rangle$.
In this case, Eqs.~(\ref{2pointav}), (\ref{3pointav}) and (\ref{4pointavc}) reduce to~\cite{Alver:2008zza}:
\begin{eqnarray}
\label{avsources}
\langle\delta f\delta g\rangle
&=&\frac{\langle\hat f\hat g\rangle}{N}\cr
\langle\delta f\delta g\delta h\rangle
&=&\frac{\langle\hat f\hat g\hat h\rangle}{N^2}\cr
\langle\delta f\delta g\delta h\delta k\rangle_c
&=&\frac{1}{N^3}\left(\langle\hat f\hat g\hat h\hat k\rangle
  -\langle\hat f\hat g\rangle\langle\hat h\hat k\rangle\right.\cr
  &&\left. -\langle\hat f\hat h\rangle\langle\hat g\hat k\rangle
  -\langle\hat f\hat k\rangle\langle\hat g\hat h\rangle\right),
\end{eqnarray}
where we have used the shorthand $\hat f\equiv f-\langle f\rangle$. 
For pointlike sources, the limit of large $N$ is also the limit of small fluctuations.
The perturbative expansion of Sec.~\ref{s:pert} is equivalent to an expansion in $1/N$~\cite{Alver:2008zza}.

For pointlike sources, Eq.~(\ref{oldresult}) simplifies to:
\begin{equation}
  \label{oldresultpointlike}
c_n\{2\}=\frac{1}{N}\frac{\langle r^{2n}\rangle}{\langle r^{n}\rangle^2},
\end{equation}
where $r\equiv|{\bf s}|$.
Adding the contributions of Eqs.~(\ref{c4A}), (\ref{c4B}) and (\ref{c4C}), one obtains: 
\begin{widetext}
  \begin{eqnarray}
    \label{pointlikeresults}
  c_n\{4\}&=&\frac{1}{N^3}\left(
  \frac{\langle r^{4n}\rangle-2\langle r^{2n}\rangle^2}{\langle r^n\rangle^4}
  -8\frac{\langle r^{3n}\rangle\langle r^{2n}\rangle}{\langle r^n\rangle^5}
  +8\frac{\langle r^{2n}\rangle^3}{\langle r^n\rangle^6}\right)
  \cr
  SC(3,2)&=&\frac{1}{N^3}\left(
  \frac{\langle r^{10}\rangle-\langle r^{4}\rangle\langle r^{6}\rangle}{\langle r^2\rangle^2\langle r^3\rangle^2}
-2 \frac{\langle r^{6}\rangle\langle r^{7}\rangle}{\langle r^2\rangle^2\langle r^3\rangle^3}
-2 \frac{\langle r^{4}\rangle\langle r^{8}\rangle}{\langle r^2\rangle^3\langle r^3\rangle^2}
  -6  \frac{\langle r^{4}\rangle\langle r^{6}\rangle}{\langle r^2\rangle^2\langle r^3\rangle^2}
  +4  \frac{\langle r^{4}\rangle\langle r^{6}\rangle\langle r^{5}\rangle}{\langle r^2\rangle^3\langle r^3\rangle^3}
  +9  \frac{\langle r^{4}\rangle^2}{\langle r^2\rangle\langle r^3\rangle^2}
 \right).
\end{eqnarray}
\end{widetext}
The expression of $c_n\{4\}$ coincides with that derived in \cite{Alver:2008zza} for $n=2$, and then extended in \cite{Bhalerao:2011bp} to $n=3$.
The result for $SC(3,2)$ is new.
These equations illustrate that these quantities depend in a non-trivial way on the initial density profile.
In particular, there are positive and negative terms, which are typically of the same order of magnitude. 
More importantly, the above results illustrate the symmetry between $c_n\{4\}$ and $SC(3,2)$.
In the next section, we introduce scaled observables which allow to compare their magnitudes. 


\section{Measures of non-Gaussianity}
\label{s:standardized}

The natural dimensionless observable is, rather than the kurtosis itself, the kurtosis divided by the square of the variance, called standardized kurtosis.
By scaling $c_n\{4\}$ and $SC(3,2)$, one obtains: 
\begin{eqnarray}
  \label{standardized}
  \frac{c_n\{4\}}{c_n\{2\}^2}
  &=& \frac{\langle \varepsilon_n \varepsilon_n\varepsilon_n^*\varepsilon_n^*\rangle}{\langle\varepsilon_n \varepsilon_n^*\rangle^2}-2,\cr
  sc(3,2)&\equiv&\frac{SC(3,2)}{c_2\{2\}c_3\{2\}}
=\frac{\langle \varepsilon_2 \varepsilon_3\varepsilon_2^*\varepsilon_3^*\rangle}
{\langle\varepsilon_2 \varepsilon_2^*\rangle\langle\varepsilon_3\varepsilon_3^*\rangle}-1.
\end{eqnarray}
One advantage of these quantities is that if $v_n$ is proportional to $\varepsilon_n$ in every event, the proportionality constant cancels out in the ratio.
Therefore, the standardized kurtosis of the distribution of $v_n$ is that of the distribution of $\varepsilon_n$, which allows one to compare directly models of initial conditions to experimental data.
Another advantage is that since they are ``standardized'', we can perform meaningful comparisons between their magnitudes.

In Sec.~\ref{s:mc}, we check the validity of the perturbative results of Sec.~\ref{s:pointlike} using Monte Carlo simulations.
In Sec.~\ref{s:expdata}, we compare several measures of primordial non-Gaussianity, and discuss the implications of existing data.

\subsection{Monte Carlo simulations}
\label{s:mc}

\begin{figure*}[t]
\centering
\includegraphics[width=\linewidth]{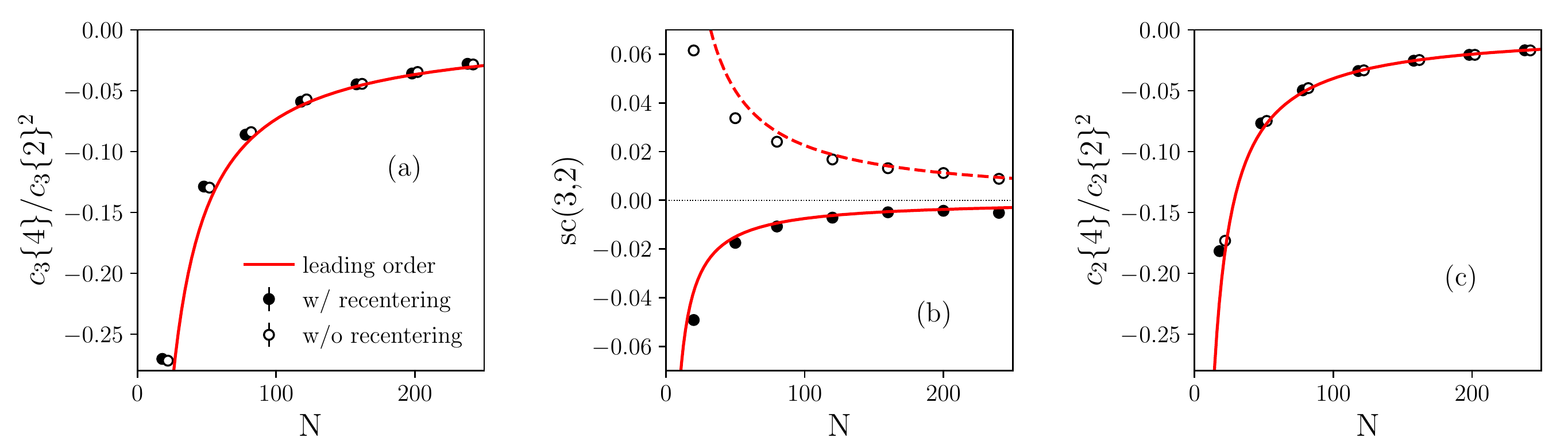} 
\caption{(Color online) 
\label{fig:gaussian}
Measures of the standardized kurtosis of the initial anisotropy for $N$ identical pointlike sources with a symmetric Gaussian distribution.
(a) Kurtosis of $\varepsilon_3$;
(b) Mixed kurtosis of $\varepsilon_2$ and $\varepsilon_3$;
(c) Kurtosis of $\varepsilon_2$. 
Symbols are Monte Carlo calculations, with recentering correction taken into account (full symbols) or switched off (open symbols).
Full lines correspond to the leading-order perturbative result, Eq.~(\ref{Gaussian}).
The dashed line in panel (b) is the perturbative result without terms due to recentering. 
In panels (a) and (c), closed and open symbols have been slightly shifted to the left and to the right, respectively, for the sake of readability. 
}
\end{figure*} 
We carry out Monte Carlo simulations using the identical source model of Sec.~\ref{s:pointlike}, in which the average density profile is a symmetric Gaussian in the transverse plane.
In this case, using Eqs.~(\ref{oldresultpointlike}) and (\ref{pointlikeresults}), one obtains
\begin{eqnarray}
  \label{Gaussian}
  \frac{c_2\{4\}}{c_2\{2\}^2}&=&-\frac{4}{N}\cr
  \frac{c_3\{4\}}{c_3\{2\}^2}&=&\frac{1}{N}\left(-\frac{69}{2}+\frac{256}{3\pi}\right)\simeq -\frac{7.34}{N}\cr
  sc(3,2)&=&-\frac{3}{4N}=-\frac{0.75}{N}. 
\end{eqnarray}
Note that the standardized kurtosises are proportional to $1/N$. 
For large $N$, fluctuations are approximately Gaussian according to the central limit theorem, which is the reason why the standardized kurtosis is small.
While the $1/N$ behaviour is generic, the dimensionless coefficient in front is sensitive to the density profile.
This sensitivity is enhanced by the fact that there are positive and negative terms of the same order of magnitude in Eq.~(\ref{pointlikeresults}):
The sum is typically smaller than any individual term. 

Figure~\ref{fig:gaussian} displays our Monte Carlo results for the standardized kurtosis, together with the analytic results (\ref{Gaussian}).
Since Eq.~(\ref{Gaussian}) is the leading-order result in $1/N$, one expects that it is valid for large $N$.
This is confirmed by the results of Fig.~\ref{fig:gaussian}, where numerical results converge to the curves as $N$ increases.
The open symbols in Fig.~\ref{fig:gaussian} are obtained by switching off the recentering correction, i.e., by setting ${\bf s}_0=0$ in Eq.~(\ref{defepsn2}).  
As expected from the perturbative calculation of Sec.~\ref{s:calcul}, this has little effect on $c_2\{4\}$ and $c_3\{4\}$.
On the other hand, it has a dramatic effect on $sc(3,2)$, which changes sign when the recentering correction is switched off.
In the perturbative calculation, the recentering correction corresponds to the fourth and sixth terms of the second line of Eq.~(\ref{pointlikeresults}), with factors $-6$ and $9$ in front.
If one omits this term, one obtains $sc(3,2)=9/(4N)$ for a Gaussian profile.
This value is displayed as a dashed line in Fig.~\ref{fig:gaussian}.
It agrees with the corresponding Monte Carlo results for large $N$, as expected.

\begin{figure*}[t]
\centering
\includegraphics[width=\linewidth]{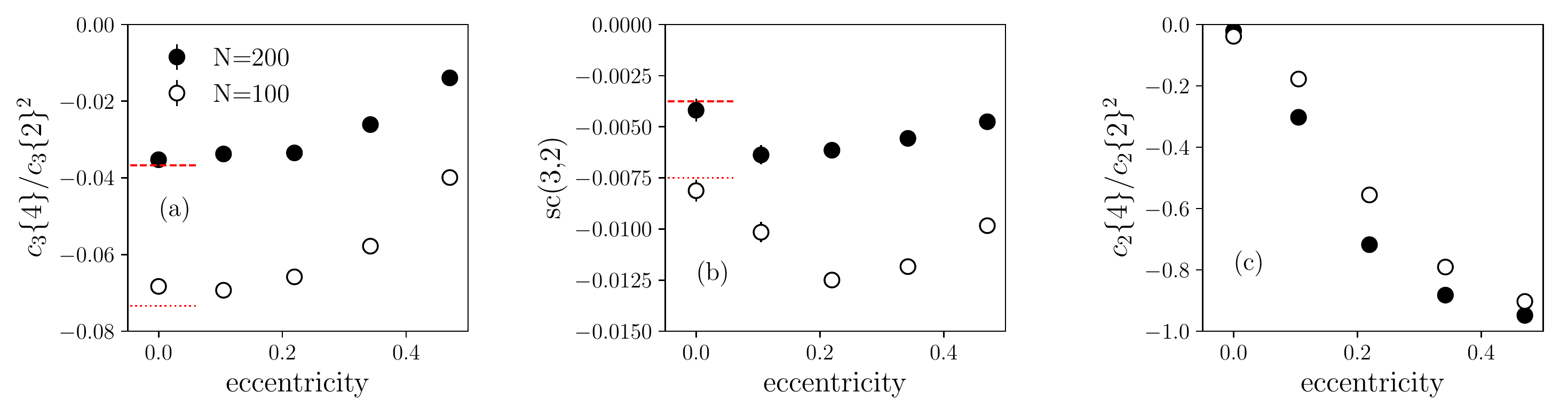} 
\caption{(Color online) 
\label{fig:eccentricity}
Same as Fig.~\ref{fig:gaussian} for an asymmetric Gaussian distribution.
Results are shown for fixed $N$ as a function of the absolute value of the mean eccentricity of the Gaussian, defined by Eq.~(\ref{bareps21}).
Recentering correction is now included everywhere.
Closed symbols: $N=200$;
Open symbols: $N=100$. 
Lines in panels (a) and (b) are the perturbative results (\ref{Gaussian}), where the eccentricity is neglected. 
}
\end{figure*} 
Our perturbative calculations are carried out for central collisions, where the mean density is azimuthally symmetric.
In order to test how results are modified for noncentral collisions, we carry out Monte Carlo calculations for an asymmetric Gaussian profile, whose widths along $x$ and $y$ differ.
We define the mean eccentricity by:~
\begin{equation}
  \label{bareps21}
  \bar\varepsilon_2\equiv\frac{\langle {\bf s}^2\rangle}{\langle |{\bf s}|^2\rangle}=\frac{\sigma_x^2-\sigma_y^2}{\sigma_x^2+\sigma_y^2}.
\end{equation}
Results are presented in Fig.~\ref{fig:eccentricity} as a function of $\bar\varepsilon_2$. 
The kurtosis of $\varepsilon_3$ (panel (a)) and the mixed kurtosis (panel (b)) depend only mildly on $\bar\varepsilon_2$.
On the other hand, the kurtosis of $\varepsilon_2$ (panel (c)) increases by orders of magnitude.
The reason is that it is driven by the mean eccentricity, not by fluctuations.
The value $-1$ corresponds to the limit where fluctuations are negligible, and $\varepsilon_2\simeq \bar\varepsilon_2$.
Therefore, the standardized kurtosis (\ref{standardized}) is not an appropriate measure of non-Gaussian fluctuations for elliptic flow.
Fluctuations of $\varepsilon_2$ in non-central collisions are discussed thoroughly in Appendix~\ref{s:noncentral}, where we show that a measure of non-Gaussian $\varepsilon_2$ fluctuations, of the same order as the standardized kurtosis, is the {\it scaled skewness\/} defined by:
\begin{equation}
\Sigma\equiv\frac{v_2\{6\}^2-v_2\{4\}^2}{v_2\{2\}^2-v_2\{4\}^2}.
\end{equation}
While this quantity is typically negative, its negative sign cannot be attributed to the positive skewness of the underlying density field, unlike the negative signs of $c_3\{4\}$ and $SC(3,2)$.

\subsection{Experimental data}
\label{s:expdata}

\begin{figure}[t]
\centering
\includegraphics[width=\linewidth]{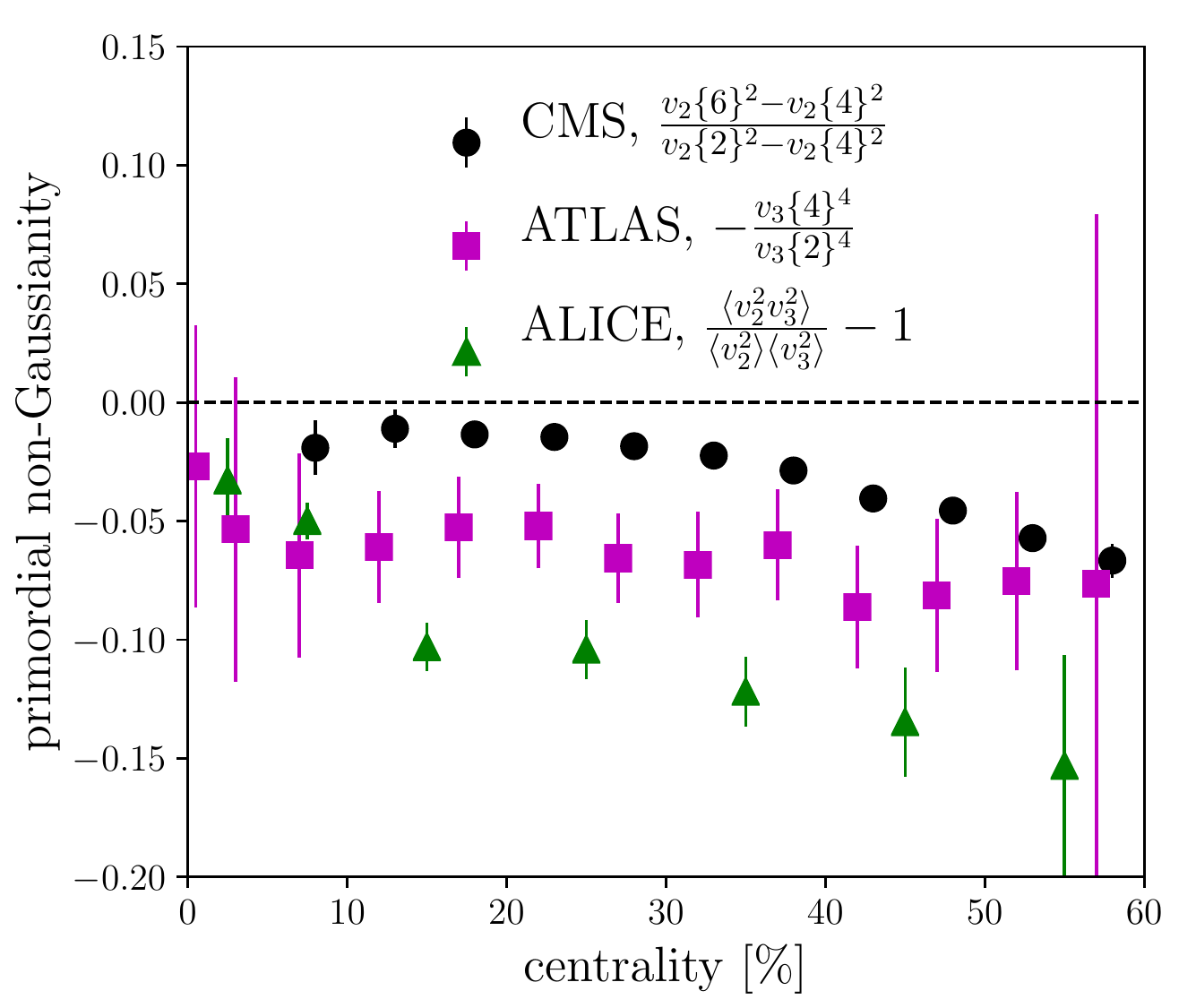} 
\caption{(Color online) 
\label{fig:data}
Three measures of non-Gaussianity as a function of the collision centrality:
scaled skewness of elliptic flow fluctuations (using CMS data~\cite{Sirunyan:2017fts});
standardized kurtosis of triangular flow (using ATLAS data \cite{Aad:2014vba});
standardized mixed kurtosis of $v_2$ and $v_3$ (using ALICE data~\cite{ALICE:2016kpq}).
}
\end{figure}

Figure~\ref{fig:data} displays three measurements of non-Gaussianity of $\varepsilon_n$ fluctuations: the scaled skewness of elliptic flow fluctuations, the kurtosis of triangular flow~\cite{Abbasi:2017ajp}, and the mixed kurtosis. 
Based on the perturbative expansion, one typically expects these quantities to be much smaller than unity, and roughly of the same order or magnitude.
Their magnitude should increase with the centrality percentile. 
Based on the results obtained for a Gaussian density profile, Eqs.~(\ref{SigmaGaussian}) and (\ref{Gaussian}), one also expects them to be negative.
These trends are confirmed by existing data to some extent. 
The scaled skewness has the expected order of magnitude and centrality dependence. 
The centrality dependence of the two standardized kurtosises is somewhat weaker than one would expect based on the $1/N$ law, taking for $N$ the number of wounded nucleons as estimated in a Glauber model~\cite{Miller:2007ri}. 
This is confirmed by Monte Carlo calculations of the kurtosis of $\varepsilon_3$~\cite{Bhalerao:2011ry,Giacalone:2017uqx} which give values in rough agreement with data, but with a stronger centrality dependence. 
The striking observation in Fig.~\ref{fig:data} is that the mixed kurtosis is larger in absolute value than the kurtosis of $v_3$, while it is a factor 10 smaller for a Gaussian density profile, Eq.~(\ref{Gaussian}).
Thus, the measured value of $sc(3,2)$ seems unusually large.
(Similar values were recently reported by the ATLAS collaboration \cite{Aaboud:2019sma}.)
This trend is confirmed if one compares with standard models of initial conditions~\cite{ALICE:2016kpq} or full hydrodynamic calculations~\cite{Gardim:2016nrrNiemi:2015qia,Acharya:2017gsw}, which predict smaller values.
However, values of $SC(3,2)$ similar to those observed in ALICE have been reported~\cite{Alba:2017hhe,Zhao:2017yhj,Moreland:2018gsh} in recent hydrodynamic calculations using the \trento{} model of initial conditions~\cite{Moreland:2014oya}.
We stress that the negative sign of the kurtosises implies that the three-point function $C_3$ of the density field is positive, as it gives the only negative contribution to the skewness (see Sec.~\ref{s:general}).

\section{Conclusions}

We have presented three different measures of non-Gaussian anisotropy fluctuations:
The scaled skewness of $v_2$ fluctuations, the kurtosis of $v_3$ fluctuations, and the mixed kurtosis between $v_2$ and $v_3$.
In hydrodynamics, these observables can be directly evaluated from the model of initial conditions, using the proportionality between $v_n$ and $\varepsilon_n$.\footnote{It has been found in extensive hydrodynamic calculations~\cite{Abbasi:2017ajp} that this proportionality breaks down, and that the kurtosis of $v_3$ is less negative than the kurtosis of $\varepsilon_3$. Such departures from linear hydrodynamic response deserve further investigations.} 
We have argued that these three quantities, which measure deviations to the central limit theorem, are typically much smaller than unity, of the same order of magnitude and with a similar centrality dependence --- their magnitude becomes larger as the centrality percentile increases. 
These trends are to some extent confirmed by existing data. 

We have carried out a perturbative expansion of anisotropies in terms of the fluctuations of the initial energy density field.
We have applied this perturbative scheme only to a few observables, but it is systematic and could be applied to any initial-state quantity, in particular participant-plane correlations~\cite{Jia:2012ju}. 

We have evaluated the kurtosises to leading order in perturbation theory. 
Even to leading order, we find that they involve the two-, three- and four-point functions of the density field, which makes their interpretation difficult.
Nonetheless, they receive a negative contributions from the three-point function only.
Hence, the observation that the kurtosises are both negative implies that the three-point function of the density field is positive, and large enough that it overcomes the other contributions.
The scaled skewness of elliptic flow fluctuations involves the two- and three-point functions of the density field, but the sign of each contribution depends on the density profile.
Therefore, it represents a less stringent probe of primordial non-Gaussianity.

\section*{Acknowledgments}

RSB would like to acknowledge the hospitality of the IPhT, Saclay,
France where a part of this work was done and the support of the CNRS
LIA (Laboratoire International Associ\'{e}) THEP (Theoretical High
Energy Physics) and the INFRE-HEPNET (IndoFrench Network on High
Energy Physics) of CEFIPRA/IFCPAR (Indo-French Center for the
Promotion of Advanced Research). RSB also acknowledges the support of
the Department of Atomic Energy, India for the award of the Raja Ramanna
Fellowship.

\appendix
\section{$\varepsilon_2$ fluctuations in non-central collisions}
\label{s:noncentral}

The overlap area of two identical nuclei in a noncentral collision is almond shaped, hence, the mean energy density profile $\langle\rho({\bf s})\rangle$ has a large elliptic deformation.
In this situation, $c_2\{4\}$ no longer represents the kurtosis of $\varepsilon_2$ fluctuations, and signatures of primordial non-Gaussianity are less easy to identify.

In this Appendix, we derive the general expression of $c_2\{4\}$ for non-central collisions.
We thus generalize the result obtained by Alver {\it et al.\/}~(Eq.~(B37) of \cite{Alver:2008zza}) to a continuous density profile.
We also provide a transparent physical interpretation of the various terms in their result. 
We then identify a signature of non-Gaussian $\varepsilon_2$ fluctuations, which we relate to experimentally-measured quantities. 

We denote by $\bar\varepsilon_2$ the value of $\varepsilon_2$ obtained by replacing $\rho({\bf s})$ with $\langle\rho({\bf s})\rangle$ in Eq.~(\ref{defepsn2}):
\begin{equation}
  \label{bareps2}
\bar\varepsilon_2\equiv\frac{\langle {\bf s}^2\rangle}{\langle |{\bf s}|^2\rangle}.
\end{equation}
We choose a coordinate frame where the impact parameter is along the $x$ axis, so that the system has $y\to -y$ symmetry and $\bar\varepsilon_2$ is real (it is negative with the convention chosen in Eq.~(\ref{bareps2})).
We decompose $\varepsilon_2$ into a (real) mean value and a (complex) fluctuation:
\begin{equation}
  \label{deceps2}
\varepsilon_2=\langle\varepsilon_2\rangle+\delta\varepsilon_2. 
\end{equation}
The magnitude of $\varepsilon_2$ fluctuations is characterized by its rms value $\sigma$:
\begin{equation}
\sigma^2\equiv\langle\delta\varepsilon_2\delta\varepsilon_2^*\rangle,
\end{equation}
whose perturbative expression was derived in~\cite{Bhalerao:2019uzw}. 
In this Appendix, we evaluate the {\it asymmetry\/} of the fluctuations, which we define by
\begin{equation}
  \label{asymmetry}
  A\equiv\langle(\delta\varepsilon_2)^2\rangle,
 \end{equation}
and the genuine non-Gaussian fluctuations, characterized by the skewness~
\begin{equation}
  \label{skewness}
S\equiv\langle(\delta\varepsilon_2)^2(\delta\varepsilon_2^*)\rangle,
\end{equation}
and the kurtosis
\begin{equation}
  \label{kurtosis}
  K\equiv   \langle(\delta\varepsilon_2)^2(\delta\varepsilon_2^*)^2\rangle-
 2 \langle(\delta\varepsilon_2)(\delta\varepsilon_2^*)\rangle^2.
\end{equation}
Inserting Eq.~(\ref{deceps2}) into Eq.~(\ref{defcn4}) and expanding in powers of $\delta\varepsilon_2$, one obtains~\cite{Bhalerao:2018anl}:
\begin{equation}
  \label{decc24}
  c_2\{4\}=-\langle\varepsilon_2\rangle^4+
  2\langle\varepsilon_2\rangle^2A
  +4\langle\varepsilon_2\rangle S
  +K.
\end{equation}
For central collisions, $\langle\varepsilon_2\rangle$ vanishes and only the kurtosis remains, so that $K$ in this equation corresponds to $c_2\{4\}$ calculated in Sec.~\ref{s:calcul}.
We consider that $K$ is not sensitive to the almond shape of the profile, and that it has the same value as for central collisions.
Note that $A$ and $S$ vanish by symmetry for central collisions.
For noncentral collisions, simplifications occur if the fluctuations are both symmetric ($A=0$) and Gaussian ($S=K=0$).
One then obtains~\cite{Voloshin:2007pc} $c_2\{4\}=-\langle\varepsilon_2\rangle^4$.

In this Appendix, we refine this result by deriving the expressions of $A$ and $S$ to leading order in perturbation theory.
In the identical source model, $A$, $S$ and $K$ are proportional to $1/N$, $1/N^2$ and $1/N^3$, respectively.
Thus, the various terms in Eq.~(\ref{decc24}) correspond to the successive orders in the $1/N$ expansion of \cite{Alver:2008zza}.\footnote{More precisely, the $1/N$ term is the sum of two contributions, where the first contribution is the asymmetry $A$, and the second contribution comes from the difference between $\langle\varepsilon_2\rangle$ and $\bar\varepsilon_2$~\cite{Bhalerao:2019uzw}.} 

We shall neglect the recentering correction to $\varepsilon_2$. 
Its contribution to the quantities we evaluate is not strictly zero for non-central collisions~\cite{Bhalerao:2019uzw}, 
but it vanishes for identical pointlike sources, and we anticipate that it is always small.
The presence of a mean eccentricity modifies  Eq.~(\ref{perteps}) to~\cite{Bhalerao:2019uzw}:
\begin{equation}
  \label{perteps2}
  \varepsilon_2=\bar\varepsilon_2+\frac{\delta {\bf s}^2}{\langle |{\bf s}|^2\rangle}
-\bar\varepsilon_2\frac{\delta {\bf ss^*}}{\langle |{\bf s}|^2\rangle}
  -\frac{(\delta {\bf ss^*})(\delta {\bf s}^2)}{\langle |{\bf s}|^2\rangle^2}
    +\bar\varepsilon_2\frac{(\delta {\bf ss^*})^2}{\langle |{\bf s}|^2\rangle^2}.
\end{equation}
There are two terms of order $\delta\rho$ and two terms of order $(\delta\rho)^2$.
Note that the average over events $\langle\varepsilon_2\rangle$ gets a contribution from second-order terms, and therefore differs from $\bar\varepsilon_2$~\cite{Bhalerao:2006tp,Bhalerao:2019uzw}.
We evaluate the asymmetry defined by Eq.~(\ref{asymmetry}), which can be rewritten as:
\begin{equation}
  \label{A2}
A=\langle (\varepsilon_2)^2\rangle -\langle \varepsilon_2\rangle^2. 
\end{equation}
We evaluate $A$ to the first non-trivial order $(\delta\rho)^2$.
Terms which are already of this order in Eq.~(\ref{perteps2}) cancel out in the difference of Eq.~(\ref{A2}). 
Therefore, only terms of order $\delta\rho$ in Eq.~(\ref{perteps2}) contribute to $A$. 
One immediately obtains
\begin{equation}
  \label{asymresult}
  A=\frac{\langle(\delta {\bf s}^2)^2\rangle-2\bar\varepsilon_2
    \langle \delta {\bf s}^2\delta {\bf ss^*}\rangle+\bar\varepsilon_2^2\langle (\delta {\bf ss^*})^2\rangle}{\langle |{\bf s}|^2\rangle^2}.
\end{equation}
The numerator is a sum of two-point averages, which is proportional to the two-point function of the density field according to Eq.~(\ref{2pointav}).
For identical pointlike sources, Eq.~(\ref{asymresult})
can be simplified along the lines of Sec.~\ref{s:pointlike}:
\begin{equation}
  \label{asymresultpointlike}
  A=\frac{\langle{\bf s}^4\rangle-2\bar\varepsilon_2
    \langle {\bf s}^3{\bf s^*}\rangle+\bar\varepsilon_2^2\langle {\bf s}^2{\bf s}^{{\bf *}2}\rangle}{N\langle |{\bf s}|^2\rangle^2}.
\end{equation}
It is typically of order $(\bar\varepsilon_2)^2/N$~\cite{Alver:2008zza}.

We now evaluate the skewness $S$, which is the connected part of the moment of order 3, to leading order in the fluctuations.
The terms of order $(\delta\rho)^3$ and $(\delta\rho)^4$ contribute to the same order after averaging over fluctuations.
We denote by $S_S$ the contribution of order $(\delta\rho)^3$ to the skewness.
It is obtained by keeping only terms of order $\delta\rho$ in Eq.~(\ref{perteps2}).
We neglect terms such as $\langle(\delta{\bf s}^2)^3\rangle$ which are of order $(\bar\varepsilon_2)^3$ and only retain the terms of order $\bar\varepsilon_2$:
\begin{equation}
\label{skewA}
  S_S=\frac{\langle(\delta {\bf s}^2)^2\delta {\bf s}^{{\bf *}2}\rangle-2\bar\varepsilon_2\langle\delta {\bf s}^2\delta {\bf ss^*}\delta {\bf s}^{{\bf *}2}\rangle}
  {\langle |{\bf s}|^2\rangle^3}.
\end{equation}
The numerator is a sum of 3-point averages, which are evaluated using Eq.~(\ref{3pointav}).
It is directly proportional to the 3-point function of the density field $C_3$,
as the contribution (\ref{c4B}) to the kurtosises. 

For identical pointlike sources, Eq.~(\ref{skewA}) gives:
\begin{equation}
\label{SSP}
  S_S=\frac{\langle{\bf s}^4{\bf s}^{{\bf *}2}\rangle-2\bar\varepsilon_2\langle|{\bf s}|^6\rangle}
 {N^2\langle |{\bf s}|^2\rangle^3}. 
\end{equation}
The two terms have opposite signs. 
For an asymmetric Gaussian density profile, they mutually cancel. 
The sign of this contribution depends on the details of the density profile, and there is no definite effect of the skewness of the density field on the skewness of $\varepsilon_2$ fluctuations.
This is at variance with the corresponding contribution to the kurtosises, Eq.~(\ref{c4B}), which is always negative.

We now evaluate the contribution of order $(\delta\rho)^4$ to the skewness, which we denote by $S_V$. 
It is obtained by taking into account only the terms of order $(\delta\rho)^2$ in one of the three $\varepsilon_2$  factors in Eq.~(\ref{skewness}), and only the terms of order $\delta\rho$ in the two remaining factors.
One is led to evaluate contributions of the type $\langle\delta_2\delta_2\delta_1\delta_1\rangle$, where we write schematically the order $(\delta\rho)^2$ term as $\delta_2\delta_2$ and the order $\delta\rho$ terms as $\delta_1$. 
This four-point average is decomposed according to Wick's theorem.
The contractions of the type $\langle\delta_2\delta_2\rangle\langle\delta_1\delta_1\rangle$ cancel out when subtracting disconnected parts.
Only the contractions of the type $\langle\delta_2\delta_1\rangle\langle\delta_2\delta_1\rangle$ remain.
One obtains:
\begin{equation}
  \label{skewB}
S_V=-4\frac{\langle\delta {\bf s}^2\delta {\bf s}^{{\bf *}2}\rangle\left(\langle\delta {\bf ss^*}\delta{\bf s}^2\rangle-\bar\varepsilon_2\langle(\delta {\bf ss^*})^2\rangle\right)}{\langle |{\bf s}|^2\rangle^4}.
\end{equation}
The numerator only contains two-point averages.
Therefore, this contribution only involves the 1- and 2-point functions of the density field, as the contribution (\ref{c4C}) to the kurtosises. 
For identical pointlike sources, this expression gives:
\begin{equation}
\label{SVP}
  S_V=
 -4\frac{\langle|{\bf s}|^4\rangle\left(\langle{\bf s}^3{\bf s^*}\rangle-\bar\varepsilon_2\langle |{\bf s}|^4\rangle\right)}{N^2\langle |{\bf s}|^2\rangle^4}.
\end{equation}
The total skewness is obtained by adding the contributions from Eqs.~(\ref{skewA}) and (\ref{skewB}):
\begin{equation}
 S=S_S+S_V.
\end{equation}
For pointlike sources, using Eqs.~(\ref{SSP}) and (\ref{SVP}), one recovers the $1/N^2$ contribution to $c_2\{4\}^4$, which was derived in~\cite{Alver:2008zza} (Eq.~(B37)).
We have seen that $S_S$ can have either sign depending on the profile. 
$S_V$ again contains two terms of opposite signs, but the first typically dominates, resulting in a negative $S_V$.
For an asymmetric Gaussian profile, one obtains
\begin{equation}
  \label{SVGaussian}
S=S_V=-\frac{8\bar\varepsilon_2}{N^2}. 
\end{equation}

\begin{figure}[t]
\begin{center}
\includegraphics[width=\linewidth]{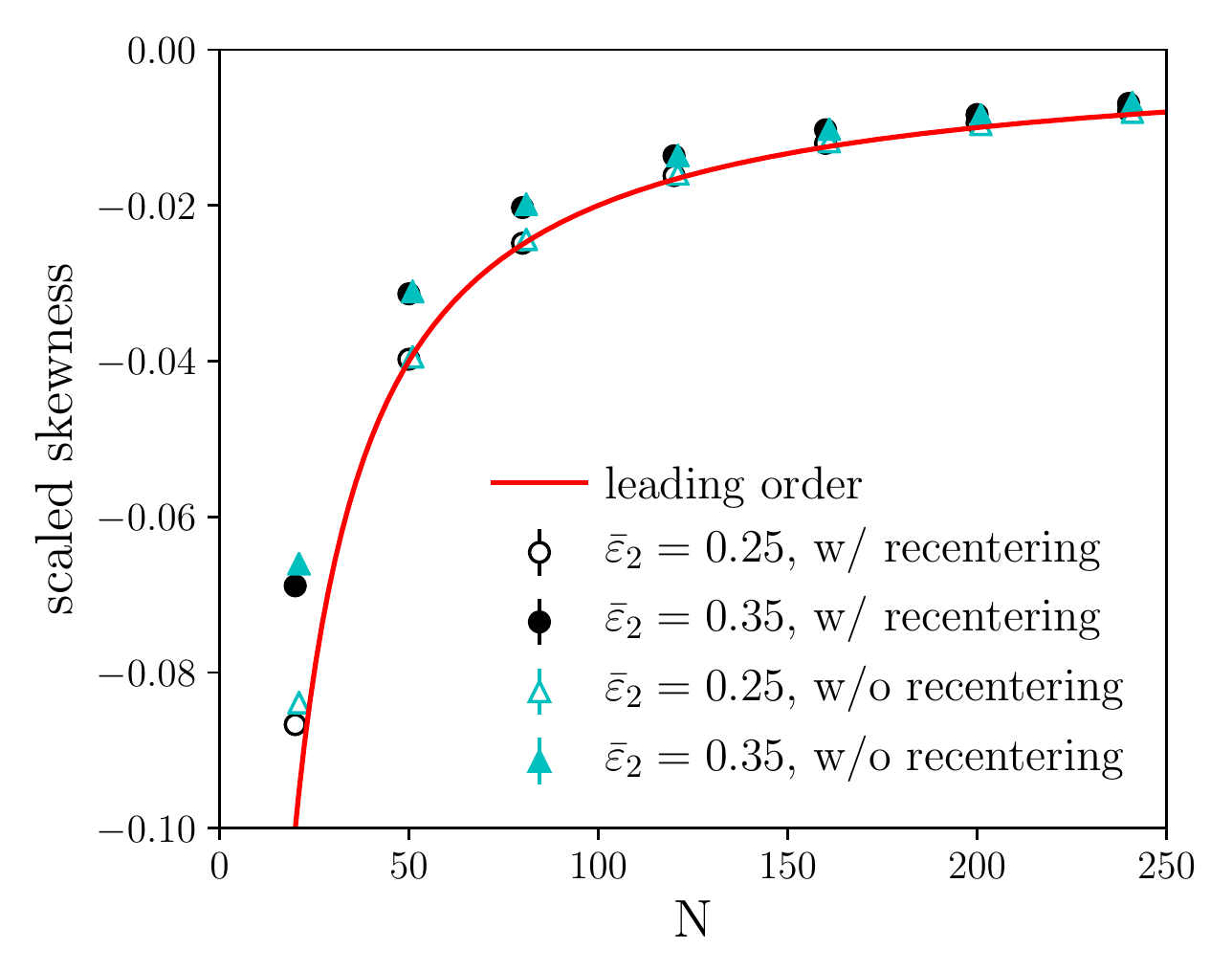} 
\end{center}
\caption{(Color online) 
\label{fig:skewness}
Scaled skewness, defined by Eq.~(\ref{defSigma}), for identical pointlike sources with an asymmetric Gaussian distribution. 
Symbols are Monte Carlo calculations with two values of $\bar\varepsilon_2$, and the line is the leading-order perturbative result (\ref{SigmaGaussian}). 
}
\end{figure} 

Eventually, we relate the skewness, which is the leading contribution to non-Gaussian $\varepsilon_2$ fluctuations in non-central collisions, to experimental observables.
We construct a dimensionless observable, $\Sigma$, which is proportional to $S$ and of the same order of magnitude as the standardized kurtosises (\ref{standardized}), i.e., $1/N$ (Eq.~(\ref{Gaussian})).
$S$ can be extracted from the small splitting between $\varepsilon_2\{6\}$ and $\varepsilon_2\{4\}$~\cite{Giacalone:2016eyu}:
\begin{equation}
  \label{64splitting}
\varepsilon_2\{6\}-\varepsilon_2\{4\}\simeq \frac{S}{4(\bar\varepsilon_2)^2},
\end{equation} 
where we have used approximate azimuthal symmetry which implies $\langle\delta\varepsilon_x^3\rangle=3\langle\delta\varepsilon_x\delta\varepsilon_y^2\rangle$. 
The quantity (\ref{64splitting}) is of order $1/(N^2\bar\varepsilon_2)$.
We next use the approximate equality $\varepsilon_2\{4\}\simeq \varepsilon_2\{6\}\simeq \bar\varepsilon_2$ to write:
\begin{equation}
\varepsilon_2\{6\}^2-\varepsilon_2\{4\}^2\simeq \frac{S}{2\bar\varepsilon_2}.
\end{equation} 
This quantity is now of order $1/N^2$.
Eventually, we make it dimensionless by dividing by the variance of $\varepsilon_2$ fluctuations $\sigma^2$:
\begin{equation}
  \label{defSigma}
\Sigma\equiv \frac{\varepsilon_2\{6\}^2-\varepsilon_2\{4\}^2}{\varepsilon_2\{2\}^2-\varepsilon_2\{4\}^2}=\frac{S}{2\bar\varepsilon_2\sigma^2}. 
\end{equation}
This quantity is of order $1/N$. If $v_2$ is proportional to $\varepsilon_2$, it can be extracted from data using
\begin{equation}
\Sigma=\frac{v_2\{6\}^2-v_2\{4\}^2}{v_2\{2\}^2-v_2\{4\}^2}.
\end{equation}
We dub this quantity the {\it scaled\/} skewness, to distinguish it from the {\it standardized\/} skewness~\cite{Giacalone:2016eyu,Sirunyan:2017fts,Acharya:2018lmh}, which is of different order, $\bar\varepsilon_2/N^{1/2}$~\cite{Bhalerao:2018anl}.
For a Gaussian density profile, $\sigma^2=2/N$. Using Eqs.~(\ref{defSigma}) and (\ref{SVGaussian}), one obtains:
\begin{equation}
  \label{SigmaGaussian}
  \Sigma=-\frac{2}{N}. 
\end{equation}
Figure~\ref{fig:skewness} displays a comparison between Monte Carlo calculations and this perturbative estimate, for two values of the mean eccentricity $\bar\varepsilon_2$.
One sees that the scaled skewness is independent of $\bar\varepsilon_2$, and has limited sensitivity to the recentering correction, which we have neglected throughout this Appendix. 
Agreement with the perturbative result (\ref{SigmaGaussian}) is good for large $N$ as expected.
It is better for the smaller value of $\bar\varepsilon_2$. 

\section{Expansion of $\varepsilon_n$ to order $(\delta\rho)^3$}
\label{s:nnlo}

Equations~(\ref{perteps}) give the expansion of Eq.~(\ref{defepsn2}) up to order $(\delta\rho)^2$.
If one includes the next order, $(\delta\rho)^3$, one obtains:
\begin{eqnarray}
  \label{nnlo}
  \varepsilon_2&=&
\frac{\delta {\bf s}^2}{\langle |{\bf s}|^2\rangle}
  -\frac{(\delta |{\bf s}|^2)(\delta {\bf s}^2)}{\langle |{\bf s}|^2\rangle^2}
  -\frac{(\delta {\bf s})^2}{\langle |{\bf s}|^2\rangle}\cr
  &&
   +\frac{(\delta |{\bf s}|^2)^2(\delta {\bf s}^2)}{\langle |{\bf s}|^2\rangle^3}
   +\frac{(\delta |{\bf s}|^2)(\delta {\bf s})^2}{\langle |{\bf s}|^2\rangle^2}
   +\frac{(\delta {\bf s}^2)(\delta {\bf s})(\delta {\bf s^*})}{\langle |{\bf s}|^2\rangle^2}
   \cr
   \varepsilon_3&=&
   \label{eq:eps3B}
\frac{\delta {\bf s}^3}{\langle |{\bf s}|^3\rangle}
  -\frac{(\delta |{\bf s}|^3)(\delta {\bf s}^3)}{\langle |{\bf s}|^3\rangle^2}
  -3\frac{(\delta {\bf s}^2)(\delta {\bf s})}{\langle |{\bf s}|^3\rangle}\cr
  &&
   +\frac{(\delta |{\bf s}|^3)^2(\delta {\bf s}^3)}{\langle |{\bf s}|^3\rangle^3}
   +3\frac{(\delta |{\bf s}|^3)(\delta {\bf s}^2)(\delta {\bf s})}{\langle |{\bf s}|^3\rangle^2}\cr
  &&
   +\frac{3}{2}\frac{(\delta {\bf s})(\delta |{\bf s}|{\bf s^*})(\delta {\bf s}^3)}{\langle |{\bf s}|^3\rangle^2}
   +\frac{3}{2}\frac{(\delta {\bf s^*})(\delta |{\bf s}|{\bf s})(\delta {\bf s}^3)}{\langle |{\bf s}|^3\rangle^2}\cr
   &&
   -\frac{9}{4}\frac{(\delta {\bf s})(\delta {\bf s^*})(\delta {\bf s}^3)\langle |{\bf s}|\rangle}{\langle |{\bf s}|^3\rangle^2}
   +2\frac{(\delta {\bf s})^3}{\langle |{\bf s}|^3\rangle}\cr
&&  +3\frac{(\delta {\bf s}^2)(\delta {\bf s})(\delta 1)}{\langle |{\bf s}|^3\rangle}.
\end{eqnarray}
Note that there are more terms of order $(\delta\rho)^3$ for $\varepsilon_3$ than for $\varepsilon_2$.
In the last line of Eq.~(\ref{eq:eps3B}), $\delta 1$ denotes the relative fluctuation of the energy, $\delta 1=\int_{\bf s}\delta\rho/\langle E\rangle$, according to the notation of Eq.~(\ref{notation}). 
This quantity appears when expanding Eq.~(\ref{defs0}) up to order $(\delta\rho)^2$. 

We now explain why the terms of order $(\delta\rho)^3$ do not contribute to the  kurtosises defined by Eq.~(\ref{defcn4}) and (\ref{defsc32}). 
When evaluating a product of four $\varepsilon_n$ factors up to order $(\delta\rho)^6$, one must include terms where one of the $\varepsilon_n$ factors is expanded up to order $(\delta\rho)^3$, while all the other factors are evaluated to leading order $\delta\rho$.
Such terms are of the form $(\delta_3\delta_3\delta_3)\delta_1\delta_1\delta_1$, and their average over events is evaluated using Eq.~(\ref{3pointav}). 
Contractions of the type $\langle\delta_3\delta_3\rangle\langle\delta_3\delta_1\rangle\langle\delta_1\delta_1\rangle$ cancel out when subtracting disconnected terms.
Only contractions of the type $\langle\delta_3\delta_1\rangle\langle\delta_3\delta_1\rangle\langle\delta_3\delta_1\rangle$ contribute to the kurtosises.
However, one easily sees that there is always one factor $\langle\delta_3\delta_1\rangle$ which vanishes by azimuthal symmetry, so that these contractions also vanish.


\begin{thebibliography}{99}

\bibitem{Chatrchyan:2012wg} 
  S.~Chatrchyan {\it et al.} [CMS Collaboration],
  Eur.\ Phys.\ J.\ C {\bf 72}, 2012 (2012)
  [arXiv:1201.3158 [nucl-ex]].



\bibitem{Aamodt:2010pa} 
  K.~Aamodt {\it et al.} [ALICE Collaboration],
  Phys.\ Rev.\ Lett.\  {\bf 105}, 252302 (2010)
  [arXiv:1011.3914 [nucl-ex]].



\bibitem{Alver:2010gr} 
  B.~Alver and G.~Roland,
  Phys.\ Rev.\ C {\bf 81}, 054905 (2010)
  Erratum: [Phys.\ Rev.\ C {\bf 82}, 039903 (2010)]
  [arXiv:1003.0194 [nucl-th]].



\bibitem{Luzum:2011mm} 
  M.~Luzum,
  J.\ Phys.\ G {\bf 38}, 124026 (2011)
  [arXiv:1107.0592 [nucl-th]].



\bibitem{Niemi:2012aj} 
  H.~Niemi, G.~S.~Denicol, H.~Holopainen and P.~Huovinen,
  Phys.\ Rev.\ C {\bf 87}, no. 5, 054901 (2013)
  [arXiv:1212.1008 [nucl-th]].



\bibitem{Teaney:2010vd} 
  D.~Teaney and L.~Yan,
  Phys.\ Rev.\ C {\bf 83}, 064904 (2011)
  [arXiv:1010.1876 [nucl-th]].



\bibitem{Alver:2008zza} 
  B.~Alver {\it et al.} [PHOBOS Collaboration],
  Phys.\ Rev.\ C {\bf 77}, 014906 (2008)
  [arXiv:0711.3724 [nucl-ex]].



\bibitem{Bhalerao:2011bp} 
  R.~S.~Bhalerao, M.~Luzum and J.~Y.~Ollitrault,
  Phys.\ Rev.\ C {\bf 84}, 054901 (2011)
  [arXiv:1107.5485 [nucl-th]].



\bibitem{Yan:2013laa} 
  L.~Yan and J.~Y.~Ollitrault,
  Phys.\ Rev.\ Lett.\  {\bf 112}, 082301 (2014)
  [arXiv:1312.6555 [nucl-th]].



\bibitem{Ade:2013ydc} 
  P.~A.~R.~Ade {\it et al.} [Planck Collaboration],
  Astron.\ Astrophys.\  {\bf 571}, A24 (2014)
  [arXiv:1303.5084 [astro-ph.CO]].



\bibitem{Bilandzic:2013kga} 
  A.~Bilandzic, C.~H.~Christensen, K.~Gulbrandsen, A.~Hansen and Y.~Zhou,
  Phys.\ Rev.\ C {\bf 89}, no. 6, 064904 (2014)
  [arXiv:1312.3572 [nucl-ex]].

\bibitem{ALICE:2016kpq} 
  J.~Adam {\it et al.} [ALICE Collaboration],
  Phys.\ Rev.\ Lett.\  {\bf 117}, 182301 (2016)
  [arXiv:1604.07663 [nucl-ex]].

\bibitem{STAR:2018fpo} 
  J.~Adam {\it et al.} [STAR Collaboration],
  Phys.\ Lett.\ B {\bf 783}, 459 (2018)
  [arXiv:1803.03876 [nucl-ex]].
  
\bibitem{Abbasi:2017ajp} 
  N.~Abbasi, D.~Allahbakhshi, A.~Davody and S.~F.~Taghavi,
  Phys.\ Rev.\ C {\bf 98}, no. 2, 024906 (2018)
  [arXiv:1704.06295 [nucl-th]].


\bibitem{ALICE:2011ab} 
  K.~Aamodt {\it et al.} [ALICE Collaboration],
  Phys.\ Rev.\ Lett.\  {\bf 107}, 032301 (2011)
  [arXiv:1105.3865 [nucl-ex]].


\bibitem{Giacalone:2016eyu} 
  G.~Giacalone, L.~Yan, J.~Noronha-Hostler and J.~Y.~Ollitrault,
  Phys.\ Rev.\ C {\bf 95}, no. 1, 014913 (2017)
  [arXiv:1608.01823 [nucl-th]].



\bibitem{Sirunyan:2017fts} 
  A.~M.~Sirunyan {\it et al.} [CMS Collaboration],
  Phys.\ Lett.\ B {\bf 789}, 643 (2019)
  [arXiv:1711.05594 [nucl-ex]].



\bibitem{Acharya:2018lmh} 
  S.~Acharya {\it et al.} [ALICE Collaboration],
  JHEP {\bf 1807}, 103 (2018)
  [arXiv:1804.02944 [nucl-ex]].



\bibitem{Blaizot:2014nia} 
  J.~P.~Blaizot, W.~Broniowski and J.~Y.~Ollitrault,
  Phys.\ Lett.\ B {\bf 738}, 166 (2014)
  [arXiv:1405.3572 [nucl-th]].



\bibitem{Gronqvist:2016hym} 
  H.~Gr\"onqvist, J.~P.~Blaizot and J.~Y.~Ollitrault,
  Phys.\ Rev.\ C {\bf 94}, no. 3, 034905 (2016)
  [arXiv:1604.07230 [nucl-th]].



\bibitem{Albacete:2017ajt} 
  J.~L.~Albacete, H.~Petersen and A.~Soto-Ontoso,
  Phys.\ Lett.\ B {\bf 778}, 128 (2018)
  [arXiv:1707.05592 [hep-ph]].


\bibitem{Sirunyan:2017uyl} 
  A.~M.~Sirunyan {\it et al.} [CMS Collaboration],
  Phys.\ Rev.\ Lett.\  {\bf 120}, no. 9, 092301 (2018)
  [arXiv:1709.09189 [nucl-ex]].

\bibitem{Aaboud:2018syf} 
  M.~Aaboud {\it et al.} [ATLAS Collaboration],
  Phys.\ Lett.\ B {\bf 789}, 444 (2019)
  [arXiv:1807.02012 [nucl-ex]].





\bibitem{Acharya:2019vdf} 
  S.~Acharya {\it et al.} [ALICE Collaboration],
  arXiv:1903.01790 [nucl-ex].



\bibitem{Alver:2006wh} 
  B.~Alver {\it et al.} [PHOBOS Collaboration],
  Phys.\ Rev.\ Lett.\  {\bf 98}, 242302 (2007)
  [nucl-ex/0610037].



\bibitem{Bhalerao:2014xra} 
  R.~S.~Bhalerao, J.~Y.~Ollitrault and S.~Pal,
  Phys.\ Lett.\ B {\bf 742}, 94 (2015)
  [arXiv:1411.5160 [nucl-th]].



\bibitem{Gardim:2011xv} 
  F.~G.~Gardim, F.~Grassi, M.~Luzum and J.~Y.~Ollitrault,
  Phys.\ Rev.\ C {\bf 85}, 024908 (2012)
  [arXiv:1111.6538 [nucl-th]].



\bibitem{Borghini:2000sa} 
  N.~Borghini, P.~M.~Dinh and J.~Y.~Ollitrault,
  Phys.\ Rev.\ C {\bf 63}, 054906 (2001)
  [nucl-th/0007063].



\bibitem{Voloshin:2007pc} 
  S.~A.~Voloshin, A.~M.~Poskanzer, A.~Tang and G.~Wang,
  Phys.\ Lett.\ B {\bf 659}, 537 (2008)
  [arXiv:0708.0800 [nucl-th]].



\bibitem{Borghini:2001vi} 
  N.~Borghini, P.~M.~Dinh and J.~Y.~Ollitrault,
  Phys.\ Rev.\ C {\bf 64}, 054901 (2001)
  [nucl-th/0105040].



\bibitem{Floerchinger:2014fta} 
  S.~Floerchinger and U.~A.~Wiedemann,
  JHEP {\bf 1408}, 005 (2014)
  [arXiv:1405.4393 [hep-ph]].



\bibitem{Giacalone:2019kgg} 
  G.~Giacalone, P.~Guerrero-Rodr\'\i guez, M.~Luzum, C.~Marquet and J.~Y.~Ollitrault,
  arXiv:1902.07168 [nucl-th].


\bibitem{Bhalerao:2006tp} 
  R.~S.~Bhalerao and J.~Y.~Ollitrault,
  Phys.\ Lett.\ B {\bf 641}, 260 (2006)
  [nucl-th/0607009].

\bibitem{Aad:2014vba} 
  G.~Aad {\it et al.} [ATLAS Collaboration],
  Eur.\ Phys.\ J.\ C {\bf 74}, no. 11, 3157 (2014)
  [arXiv:1408.4342 [hep-ex]].


\bibitem{Miller:2007ri} 
  M.~L.~Miller, K.~Reygers, S.~J.~Sanders and P.~Steinberg,
  Ann.\ Rev.\ Nucl.\ Part.\ Sci.\  {\bf 57}, 205 (2007)
  [nucl-ex/0701025].



\bibitem{Bhalerao:2011ry} 
  R.~S.~Bhalerao, M.~Luzum and J.~Y.~Ollitrault,
  J.\ Phys.\ G {\bf 38}, 124055 (2011)
  [arXiv:1106.4940 [nucl-ex]].



\bibitem{Giacalone:2017uqx} 
  G.~Giacalone, J.~Noronha-Hostler and J.~Y.~Ollitrault,
  Phys.\ Rev.\ C {\bf 95}, no. 5, 054910 (2017)
  [arXiv:1702.01730 [nucl-th]].

\bibitem{Aaboud:2019sma} 
  M.~Aaboud {\it et al.} [ATLAS Collaboration],
  arXiv:1904.04808 [nucl-ex].

\bibitem{Gardim:2016nrr} 
  F.~G.~Gardim, F.~Grassi, M.~Luzum and J.~Noronha-Hostler,
  Phys.\ Rev.\ C {\bf 95}, no. 3, 034901 (2017)
  [arXiv:1608.02982 [nucl-th]].
  
\bibitem{Niemi:2015qia} 
  H.~Niemi, K.~J.~Eskola and R.~Paatelainen,
  Phys.\ Rev.\ C {\bf 93}, no. 2, 024907 (2016)
  [arXiv:1505.02677 [hep-ph]].

\bibitem{Acharya:2017gsw} 
  S.~Acharya {\it et al.} [ALICE Collaboration],
  Phys.\ Rev.\ C {\bf 97}, no. 2, 024906 (2018)
  [arXiv:1709.01127 [nucl-ex]].

  

\bibitem{Alba:2017hhe} 
  P.~Alba, V.~Mantovani Sarti, J.~Noronha, J.~Noronha-Hostler, P.~Parotto, I.~Portillo Vazquez and C.~Ratti,
  Phys.\ Rev.\ C {\bf 98}, no. 3, 034909 (2018)
  [arXiv:1711.05207 [nucl-th]].

\bibitem{Zhao:2017yhj} 
  W.~Zhao, H.~j.~Xu and H.~Song,
  Eur.\ Phys.\ J.\ C {\bf 77}, no. 9, 645 (2017)
  [arXiv:1703.10792 [nucl-th]].

\bibitem{Moreland:2018gsh} 
  J.~S.~Moreland, J.~E.~Bernhard and S.~A.~Bass,
  arXiv:1808.02106 [nucl-th].

\bibitem{Moreland:2014oya} 
  J.~S.~Moreland, J.~E.~Bernhard and S.~A.~Bass,
  Phys.\ Rev.\ C {\bf 92}, no. 1, 011901 (2015)
  [arXiv:1412.4708 [nucl-th]].

\bibitem{Jia:2012ju} 
  J.~Jia and D.~Teaney,
  Eur.\ Phys.\ J.\ C {\bf 73}, 2558 (2013)
  [arXiv:1205.3585 [nucl-ex]].


\bibitem{Bhalerao:2019uzw} 
  R.~S.~Bhalerao, G.~Giacalone, P.~Guerrero-Rodr\'\i guez, M.~Luzum, C.~Marquet and J.~Y.~Ollitrault,
  arXiv:1903.06366 [nucl-th].
  
\bibitem{Bhalerao:2018anl} 
  R.~S.~Bhalerao, G.~Giacalone and J.~Y.~Ollitrault,
  Phys.\ Rev.\ C {\bf 99}, no. 1, 014907 (2019)
  [arXiv:1811.00837 [nucl-th]].


\end{thebibliography}
\end{document}